\documentclass[apjl]{emulateapj}

\usepackage{graphicx}
\usepackage{enumerate}
\usepackage{lipsum}
\usepackage{mathrsfs}
\usepackage{amsmath}
\usepackage{amssymb}
\usepackage[figuresright]{rotating}
\usepackage{bm}

\makeatletter

\newcommand{\Rmnum}[1]{\expandafter\@slowromancap\romannumeral #1@}
\makeatother

\def\hi{H{\scriptsize \Rmnum{1}}}
\def\hii{H{\scriptsize \Rmnum{1}\Rmnum{1}}}
\def\ci{C{\scriptsize \Rmnum{1}}}

\def\cms{cm$^{-2}$}
\def\cm3{cm$^{-3}$}

\def\kms{km~s$^{-1}$}

\def\nh3{NH$_3$}
\def\n2h{N$_2$H$^+$}
\def\co{$^{12}$CO}
\def\13co{$^{13}$CO}
\def\c18o{C$^{18}$O}
\def\hc3n{HC$_3$N}
\def\h2{H$_2$}
\def\nh{n(H$_2$)}

\def\lp{\>\> .}
\def\lc{\>\> ,}

\shorttitle{Evolution of OH and CO-dark Molecular Gas Fraction Across a Molecular Cloud Boundary in Taurus}
\shortauthors{Duo Xu. et al.}

\slugcomment{}

\begin{document}

\title{ {Evolution of OH and CO-dark Molecular Gas Fraction Across a Molecular Cloud Boundary in Taurus} }

\author{Duo Xu\altaffilmark{1{,2}}{*} , Di Li\altaffilmark{1{,3}}{*} , Nannan Yue\altaffilmark{1{,2}} , Paul F. Goldsmith\altaffilmark{4} }
\affil{
$^1$ National Astronomical Observatories, Chinese Academy of Sciences, A20 Datun Road, Chaoyang District, Beijing 100012, China\\
{$^2$ University of Chinese Academy of Sciences, Beijing 100049, China\\}
{$^3$ Key Laboratory for Radio Astronomy, Chinese Academy of Sciences\\}
$^4$ Jet Propulsion Laboratory, California Institute of Technology, Pasadena, CA 91109, USA
}

\email{{*}Email: xuduo117@nao.cas.cn, dili@nao.cas.cn}

\begin{abstract}
We present observations of \co~{J=1-0}, \13co~{J=1-0}, \hi, and all four ground-state transitions of the hydroxyl (OH) radical toward a sharp boundary region of the Taurus molecular cloud. Based on a PDR model that reproduces
CO and [\ci] emission from the same region, we {modeled} the three OH transitions, 1612, 1665, 1667 {MHz} successfully through {escape probability non-LTE radiative transfer model} calculations. We {could not} reproduce the 1720 {MHz} observations, {due to un-modeled pumping mechanisms, of which the most likely candidate is a C-shock}. The abundance of OH and {CO-dark molecular gas~(DMG)} are well constrained. The OH abundance [OH]/[\h2] decreases from 8$\times 10^{-7}$ to 1$\times 10^{-7}$ as $A_v$ increases from 0.4 to 2.7 mag, following an empirical law 
\begin{center} 
[OH]/[\h2]$=1.5\times10^{-7}+9.0\times10^{-7}\times \exp(-A_{v}/0.81)$, 
\end{center}
{which is higher than PDR model predictions for low extinction regions by a factor of 80. The overabundance of OH at extinctions at or below 1 mag is likely the result of a C-shock.} The {dark gas fraction (DGF, defined }as fraction of {molecular} gas without detectable CO emission) decreases from 80\% to 20\%, following a gaussian profile 
\begin{center}
DGF$=0.90\times \exp(-(\frac{A_{v}-0.79}{0.71})^2)$.\\
\end{center}
This trend of the DGF is consistent with {our understanding that the} DGF drops at low visual extinction due to photodissociation of \h2\ {and} drops at high visual extinction due to CO formation. The DGF peaks in the extinction range where \h2 has already formed and achieved self-shielding but \co ~has not. Two narrow velocity components with a peak-to-peak spacing of $\sim$ 1 {km~s$^{-1}$} were clearly identified. Their relative intensity and variation in space and frequency suggest colliding streams or gas flows at the boundary region. 

\end{abstract}


\section{Introduction}

The formation of molecular hydrogen is a critical step in the transformation of interstellar gas into new stars. Various complex processes in the transformation between atomic gas and molecular gas result in changes of the physical state of gas. These changes affect the star formation process in galaxies, and thus their evolution. It is of great significance to study the physical conditions in regions where the transformation between atomic gases and molecular gases occurs. The boundaries of molecular clouds are locations where the chemical composition changes considerably and primarily atomic gas transforms into molecular gas. 

Apart from traditional tracers such as CO and \hi, the hydroxyl radical, OH, is thought to be an excellent tracer to determine the physical conditions of the early state ISM, where CO is absent (Li et al. 2015). Such early state molecular ISM appears in boundary regions. A three-way comparison of \hi, OH and CO lines is expected to show molecular features in OH which are not traced by CO, but which highlight the transition from atomic gas (seen in the \hi ~line) to molecular gas. 

Additionally, OH can be the tracer of ``J-shocks" (Medling et al. 2015) and ``C-shocks" (Anderl et al. 2013). Shock waves play a major role in the ISM (Timmermann 1998). When shock waves propagate through the molecular ISM the ambient gas is compressed, heated, and accelerated. Furthermore, the composition of the gas is significantly changed when chemical reactions occur especially in the warm shocked gas located in outermost parts of the transition region. {The abundance of OH can vary from $10^{-5}$ to $10^{-11}$ when a 10 \kms~C-type shock propagates into a diffuse cloud with n$_{\rm {H}}$=50 cm$^{-3}$ (Draine \& Katz 1986).} Thus, searching for the evidence of shocks at the boundary region via OH lines is of great significance. 

The boundary of molecular clouds is the region where all the physical and chemical processes mentioned above take place. As the dramatic changes of species at the boundaries, the physical conditions at the boundaries are distinct from those in other places. A clear example of cloud boundaries can be found in Taurus (Goldsmith et al.\ 2008) north east of the TMC1 region. Orr et al.\ (2014) compared the observations of Taurus boundaries with line intensities produced by the Meudon PDR code. They found a low ratio of $^{12}$C to $^{13}$C $\sim$ 43, and a highly depleted sulfur abundance (by a factor $\ge50$) to explain the very low [\ci] emission. Moreover, Goldsmith et al.\ (2010) found an unexpectedly high degree of excitation of the \h2\ in the boundary layer of Taurus molecular cloud. They believed that an enhanced heating rate may be the result of turbulent dissipation.

As we are interested in the changes of chemical composition across the boundaries to study the transition between atomic- and molecular-dominated gas,we need to find a boundary of molecular clouds without significant UV enhancement. \h2 can be destroyed by UV photons, making the statistical equilibrium function complex. Moreover, it is easier to detect the emission of molecular tracers such as OH without an enhanced UV field. We thus observed the Taurus boundary studied by Goldsmith et al.\ (2010) and Orr et al.\ (2014), which has a relatively low UV field {between $\chi$=0.3 and 0.8 in units of the Draine's field} (Flagey et al. 2009; Pineda et al. 2010) and little foreground or background visual extinction (Padoan et al. 2002), making it favorable for the comparison of observations with physical models.

We have carried out observations of the Taurus boundary in four OH transitions (1612, 1665, 1667, and 1720 MHz) using the 305 m Arecibo Telescope. We made a total of five cuts across the boundary region each with 17 pointings 3 arcminute apart (Fig.~1). We describe the observations of OH across the boundary region, and the \hi, \co\ {J=1-0}, and $^{13}$CO {J=1-0} map of Taurus molecular cloud in Section~2. We analyze the OH spectrum and derive the physical parameters of \13co~across the boundary in Section~3. We use a cylindrical model and RADEX to fit OH lines to determine the physical parameters in Section~4. We discusse the conjugate emission of OH and pumping mechanisms in Section~5. In Section~6, we summarize our results and conclusions from this study.

\section{Observations and Data}
\label{Observations}
We carried out observations of OH with the Arecibo Telescope in Project a2813. We extracted \hi~data from the GALFA-\hi~survey, and extracted \co~{J=1-0} and \13co~{J=1-0} data from the FCRAO Taurus survey.
\subsection{OH Observations}
The OH observations were taken using the L-band wide receiver (with frequency range 1.55-1.82 GHz) on October 28-31, 2013. We observed four OH transition lines at the rest frequencies of 1612.231, 1665.402, 1667.359, and 1720.530 MHz with the total power ON mode. Spectra were obtained with the Arecibo WAPP correlator with nine-level sampling and 4096 spectral channels for each line in each polarization. The spectral bandwidth was 3.13 MHz for a channel spacing of about 763 Hz, or 0.142 km s$^{-1}$. The average system temperature was about 31 K. The main beam of the antenna pattern had a full-width at half-maximum (FWHM) beam-width of 3$'$. Spectra were taken at 17$\times$5 positions across the {Taurus boundary region (TBR)}, as seen in Fig.~1. An integration time of 300 seconds per position was used resulting in an RMS noise level of about 0.027 K. 

\begin{figure*}[htp]
\includegraphics[width=1.00\linewidth]{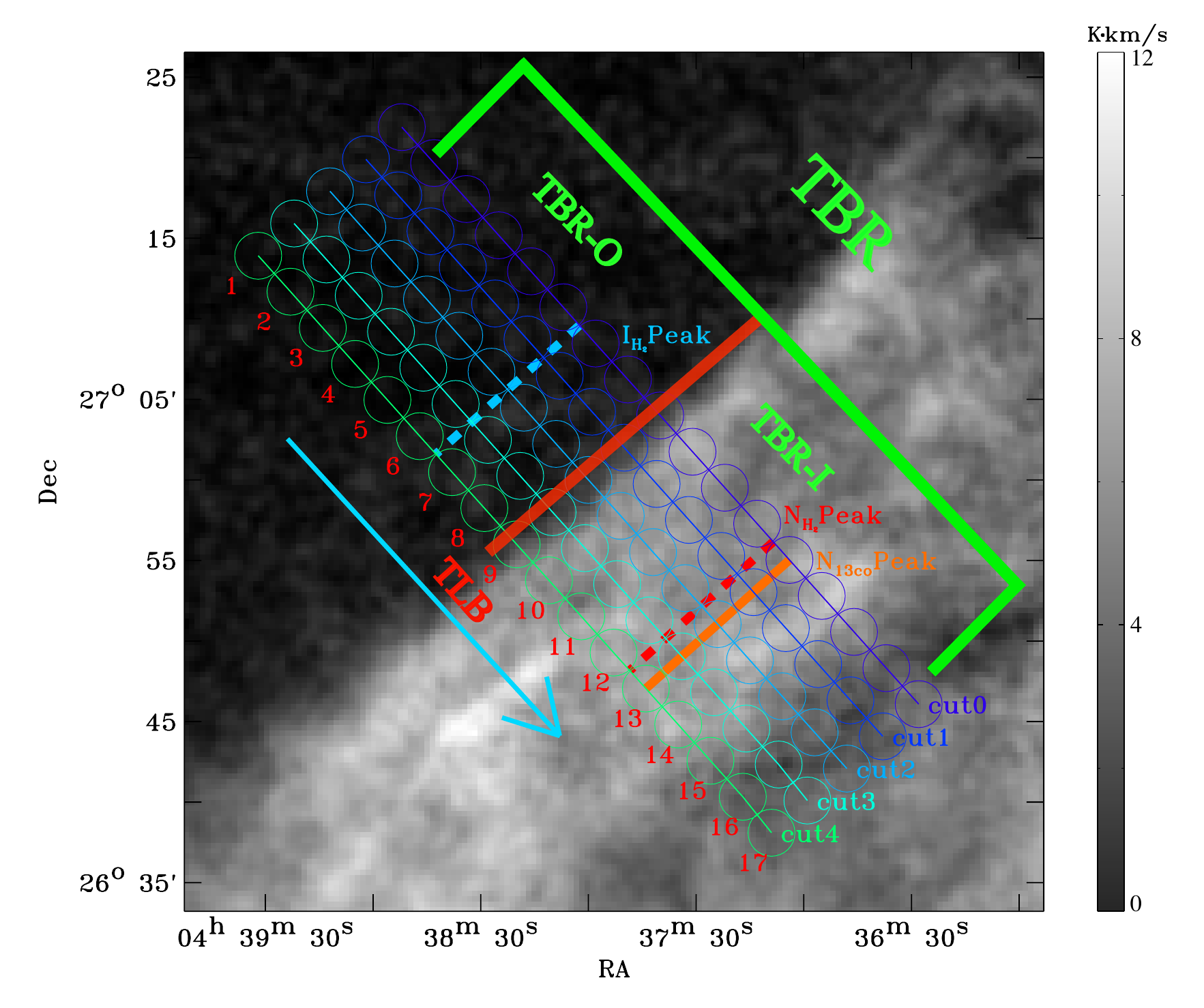}
\caption{The {boundary} region in \13co\ {J=1-0} peak intensity, with observed positions indicated. The size of circle indicates the OH beam size $\sim$ 3$^{\prime}$. The numbers of positions are shown in the figure. The whole region is Taurus {boundary} region and  is denoted {TBR}. The Taurus linear {boundary (TLB)} located at position 9 is shown as red line. The outside and inside region of the {TBR} are abbreviated as {TBR-O and TBR-I} respectively. The peak intensity of {two lowest rotational transitions of \h2, S(0) and S(1),} is located between position 6 and position 7 {(Goldsmith et al. 2010)}. The peak column density of \h2\ is located between position 12 and position 13. The peak column density of \13co\ is located at position 13. The arrow in the figure indicates the direction we present spectral line maps.  }
\label{fig.cut_ra_dec}
\end{figure*}

\subsection{\hi\ Data}
\hi\ 21 cm observations of the Taurus {boundary} region were extracted from the results of the Galactic Arecibo L-Band Feed Array \hi\ (GALFA-\hi\ ) survey (Peek et al. 2011). GALFA-\hi\ uses ALFA, a seven-beam array of receivers mounted at the focal plane of the 305 m Arecibo telescope, to map \hi\ emission in the Galaxy. {GALFA-\hi\ is a  survey of the Galactic interstellar medium in the 21 cm line hyperfine transition of neutral hydrogen which covers a large-area (13,000 deg$^{2}$) with $\sim $ 4$^{\prime}$ resolution, and has high spectral resolution (0.18 km s$^{-1}$) with broad velocity coverage ($-700$ km s$^{-1}$  $<$ v$\rm_{LSR}$$ <$ $+700$ km s$^{-1}$).} Typical RMS noise is 80 mK in a 1 km s$^{-1}$ channel.

\subsection{\co\ and \13co\ Data}

The \co\ {J=1-0} and $^{13}$CO {J=1-0} observations were taken simultaneously between  2003 and 2005 using the 13.7 m Five College Radio Astronomy Observatory (FCRAO) Telescope~(Narayanan et al. 2008).  The map is centered at $\alpha(2000.0)=04^h 32^m 44.6^s$, $\delta(2000.0)=24^\circ 25' 13.08''$, with an area of $\sim 98\ \rm deg^2$. The main beam of the antenna pattern had a full-width to half-maximum (FWHM) beam-width of 45$''$ for $^{12}$CO, and 47$''$ for $^{13}$CO. The angular spacing (pixel size) of the resampled on the fly (OTF) data is 20'' (Goldsmith et al. 2008), which corresponds to a physical scale of $\approx 0.014\rm\ pc$ at a distance of $D=140\ {\rm pc}$. The data have a mean rms antenna temperature of 0.28 K and 0.125 K, in channels of 0.26 km s$^{-1}$ and 0.27 km s$^{-1}$ width for $^{12}$CO and $^{13}$CO, respectively.

\section{Analysis}
\label{Analysis}

\subsection{Spectral Analysis}
\label{Spectrum Analysis}
The locations of the positions for the telescope pointing used to study the {TBR} are shown in Fig.~1. To examine the transition zone with higher signal to noise ratio, we averaged all five cuts of spectra of OH 1612 MHz, 1665 MHz, 1667 MHz, 1720 MHz, \co~{J=1-0}, \13co~{J=1-0}, and \hi, as shown in Fig.~2. {The \co~{J=1-0} and \13co~{J=1-0} spectra were convolved to the OH beam size of 3$^{\prime}$ at each position.} The emission lines of OH, \co~{J=1-0}, \13co~{J=1-0}, and the absorption lines of \hi\ are well matched in velocity. Especially, the emission lines of OH 1665 MHz at positions 10 to 12 all have two components, and the spectral of \hi\ at the same point have two corresponding narrow absorption components, as shown in Fig.~3.

We did gaussian fitting of the OH 1612 MHz, 1665 MHz, 1667 MHz, 1720 MHz spectral with two gaussian components, the fitting of \co ~ and \13co spectra with a single gaussian component, and the fitting of \hi\ spectra with three gaussian components. We show the spectra and the fitted profiles in Fig.~2. The line ratio between OH 1665 MHz and 1667 MHz is greater than 1 in the outside {TBR region (TBR-O)} as shown in Fig.~2. Under the assumption of LTE, the line ratio between 1665 MHz and 1667 MHz ranges from 0.6 to 1. The line ratio greater than 1 indicates a deviation from LTE for OH.

\begin{figure*}[htp]
\centering
\includegraphics[width=0.90\linewidth]{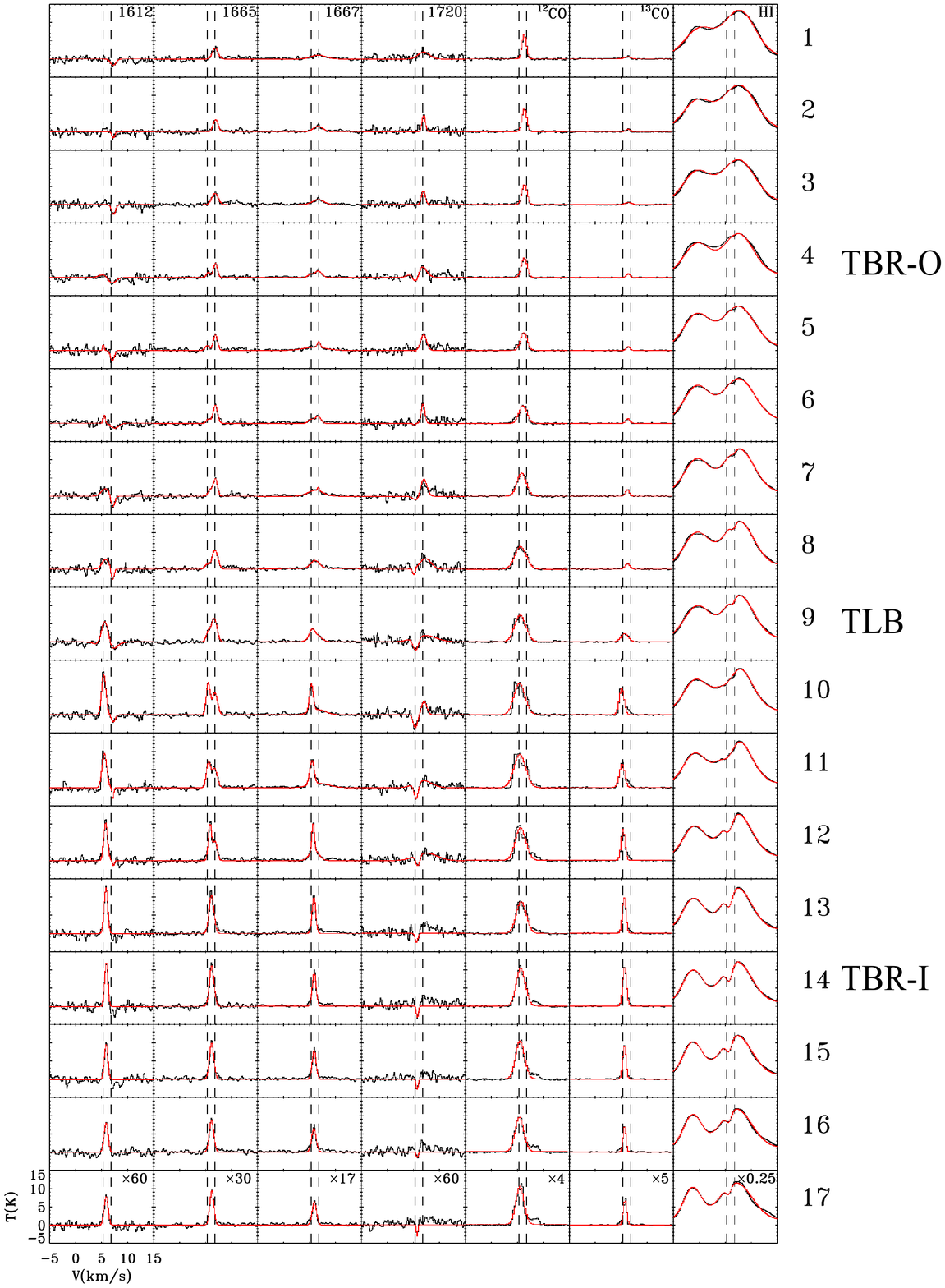}
\vspace{-10mm}
\caption{Average spectra of all five cuts of OH 1612 MHz, 1665 MHz, 1667 MHz, 1720 MHz, \co\ {J=1-0}, \13co\ {J=1-0}, and \hi\ overlaid with corresponding fitted gaussian profile(red curve). {The \co~{J=1-0} and \13co~{J=1-0} spectra were convolved to the OH beam size of 3$^{\prime}$ at each position.} We fitted the OH 1612 MHz, 1665 MHz, 1667 MHz, 1720 MHz spectra with two gaussian components, fitted the \co~{J=1-0} and \13co~{J=1-0} spectra with single gaussian component, and fitted the \hi\ spectra with three gaussian components. The vertical dashed lines indicate the central velocities of the two components of OH 1665 MHz at position 10.}
\label{fig.add_oh_co_hi_gauss}
\end{figure*}

\begin{figure*}[htp]
\includegraphics[width=1.0\linewidth]{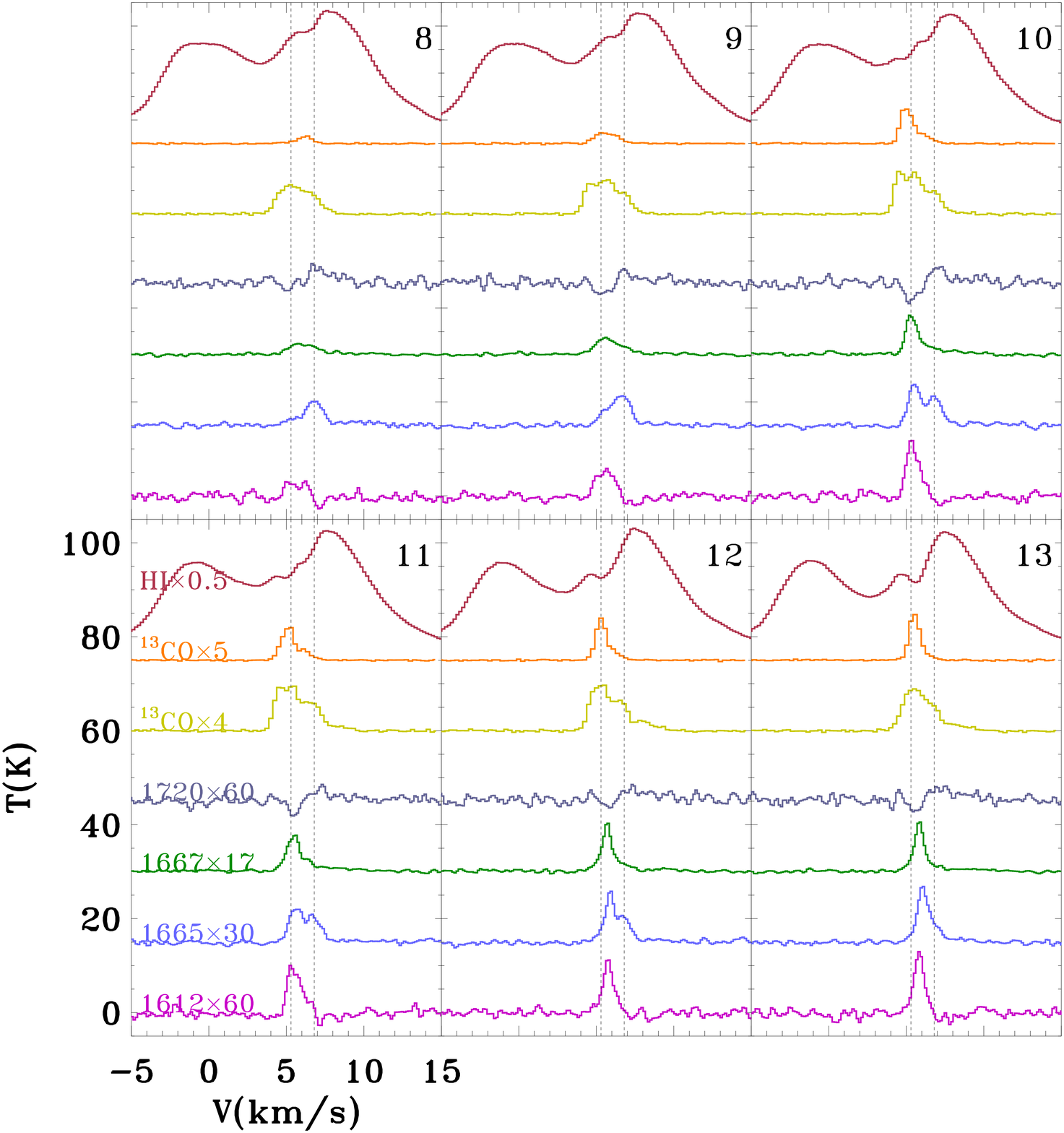}
\vspace{-10mm}
\caption{Average spectra of all five cuts of OH 1612 MHz, 1665 MHz, 1667 MHz, 1720 MHz, \co~{J=1-0}, \13co~{J=1-0}, and \hi\ at positions 8 to 13. {The \co~{J=1-0} and \13co~{J=1-0} spectra were convolved to the OH beam size of 3$^{\prime}$ at each position.} The vertical dashed lines indicate the central velocities of the two components of OH 1665 MHz at position 10.}
\label{fig.add_oh_co_hi_10_13}
\end{figure*}

We show the change of peak intensity of two components in OH 1665 MHz and a single component in \13co across the {TBR} in Fig.~4. Component 1 of OH 1665 MHz spectrum at 5.3 \kms~appears after position 4, and gets stronger across the {TBR}. Component 2 at 6.8 \kms~appears in the {TBR-O} and gets faint in the inside {TBR region (TBR-I)}, and disappears after position 13. The central velocities of \13co shift from  6.3 \kms~to 5.7 \kms. We assume that each component of the OH emission indicates a gas stream. When the intensity of OH component 2 is stronger than that of component 1 in {TBR-O}, the central velocity of \13co~is located at 6.3 \kms. When the intensity of OH component 1 is stronger than that of component 2 in {TBR-I}, the central velocity of \13co~is located at 5.7 \kms. The central velocity of \13co~shifts in the same way as the central velocity of stronger intensity OH component shifts. {From Fig.~3 we see that component 2 of OH 1665 gradually becomes fainter, and disappears at position 13. At the same time the central velocity of component 1 gradually shifts from 5.3 \kms\ at position 9 to 5.8 \kms\ at position 13, which indicates that the collision of two streams results in the final central velocity being located between the velocities of the two components. The central velocity of the final combined stream is located closer to component 1, which has a stronger emission line in {TBR-I}. This is consistent with the assumption of different amounts of \13co\ emission at different velocities. Not only OH 1665 MHz but also \13co\ shows the central velocity shifting after the streams collide from position 10 to position 13 in Fig.~3.}

\begin{figure*}[htp]
\includegraphics[width=1.00\linewidth]{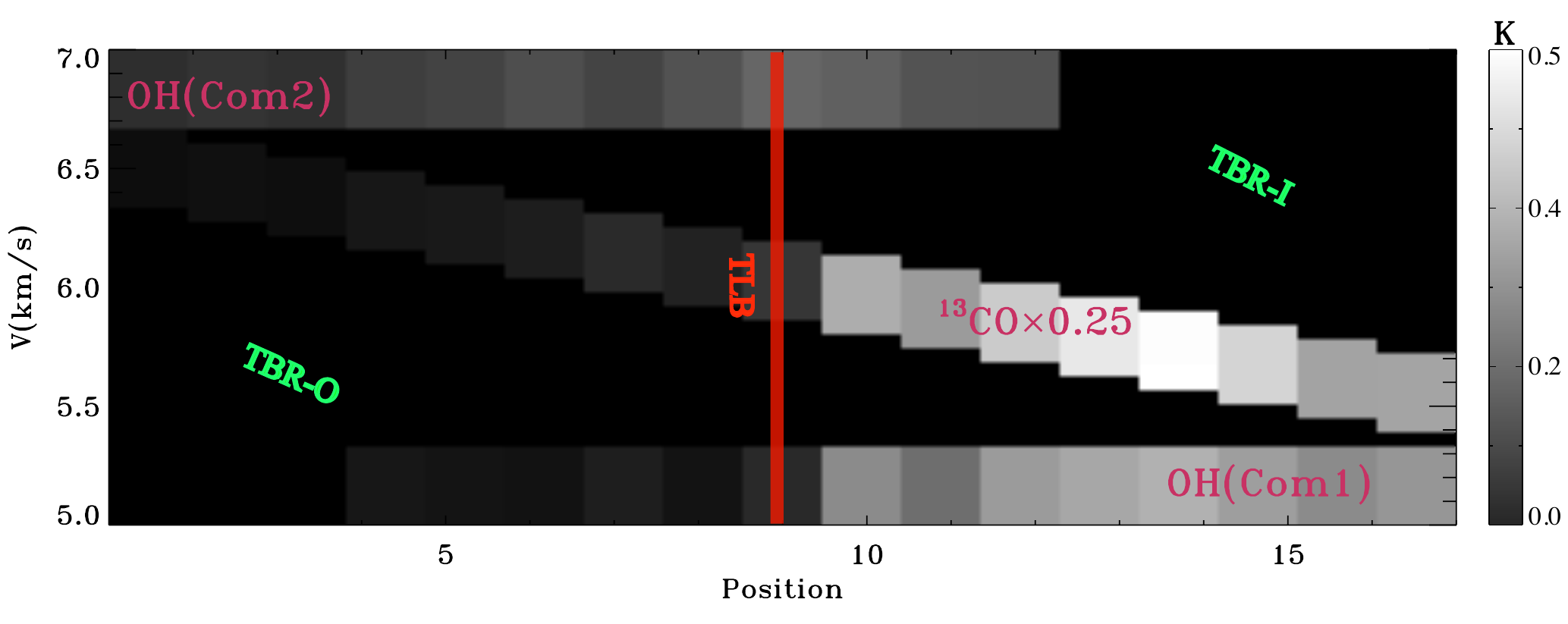}
\caption{Change of peak intensity of the two components of OH 1665 and the single component of \13co across the {TBR}. The upper stripe indicates the intensity of component 2 in OH 1665 MHz across the {TBR}. The middle oblique stripe indicates the intensity of \13co. The bottom stripe indicates the intensity of component 1 in OH 1665 MHz across the {TBR}. {The color indicates the value of peak intensity of each spectrum at each position. The peak intensity of \13co~ shown in the oblique stripe is one fourth of the observed peak intensity.} A more detailed discussion is in Section~3.1.}
\label{fig.intensity}
\end{figure*}

We show the change of line width along the cut direction in Fig.~5. We can clearly see that the line width of \co~and \13co peak at the {TLB}. This mainly due to the two gas streams which contribute to the line width of \co~and \13co. The line width of OH 1667 MHz is much wider than that of other lines in {TBR-O}. This is mainly caused by the weak emission lines of OH 1667 MHz in {TBR-O}. Since the height of OH 1667 MHz is small in {TBR-O}, the fitted gaussian profile tends to be flatter than that at other positions. So the line width of OH 1667 MHz is two times wider than that of the other three lines in {TBR-O}. Component 2 of OH 1667 MHz whose central velocity is at 6.8 \kms~is very weak in some positions in {TBR-I}, which leads to a wider line wing of the component 1, as shown in Fig.~2. The width of component 2 in {TBR-I} is wider than that of other three lines in {TBR-I}, which leads to the average line width of OH 1667 behaving in an erratic way. The line width of OH 1665 MHz is nearly constant across the {TBR}. 

The line width of \co\ is significantly greater than that of \13co. Opacity may be one of the reasons. Owing to the different abundance of \co\ and \13co, $\tau_{12}$ is almost 70 times larger than $\tau_{13}$. When  $\tau_{12}$ is much larger than unity, the term 1-exp[$-\tau_{12}\phi(\Delta \mathbf v)$] in the line profile of \co\ is much wider than that of \13co, where $\phi(\Delta \mathbf v)$ is a gaussian profile of the velocity offset from line center $\Delta \mathbf v$. We made an estimate of the line width ratio between \co\ and \13co only considering the opacity. In TBR-O, the optical depth range from 0.05 to 0.4. We took position 8 having $\tau_{^{{13}CO}}=0.2$ as an example. The line width ratio between \co\ and \13co is 1.4, which is much less than the observed ratio 2.3. Other broadening mechanisms must occur in the TBR. Park \& Hong (1995) took different broadening mechanisms such as micro-turbulence and macro-turbulence into consideration and found that the line width ratio between \co\ and \13co can range from 1 to 3 depending on the physical condition of the gas, which just corresponds to the line width ratio between \co\ and \13co\ in our work. The wider line width of \co\ may be the result of larger turbulence due to the more extended distribution of \co. According to Larson's law (Larson 1981), a larger scale of a molecular cloud leads to a larger velocity dispersion. Considering current geometrical model of photo-dissociation region (Tielens 2005), the larger self-shielding threshold of \13co\ makes it more constrained to the inner region of the boundary, which yields a narrower line width than that of \co.

\begin{figure}[htp]
\centering 
\includegraphics[width=0.5\textwidth]{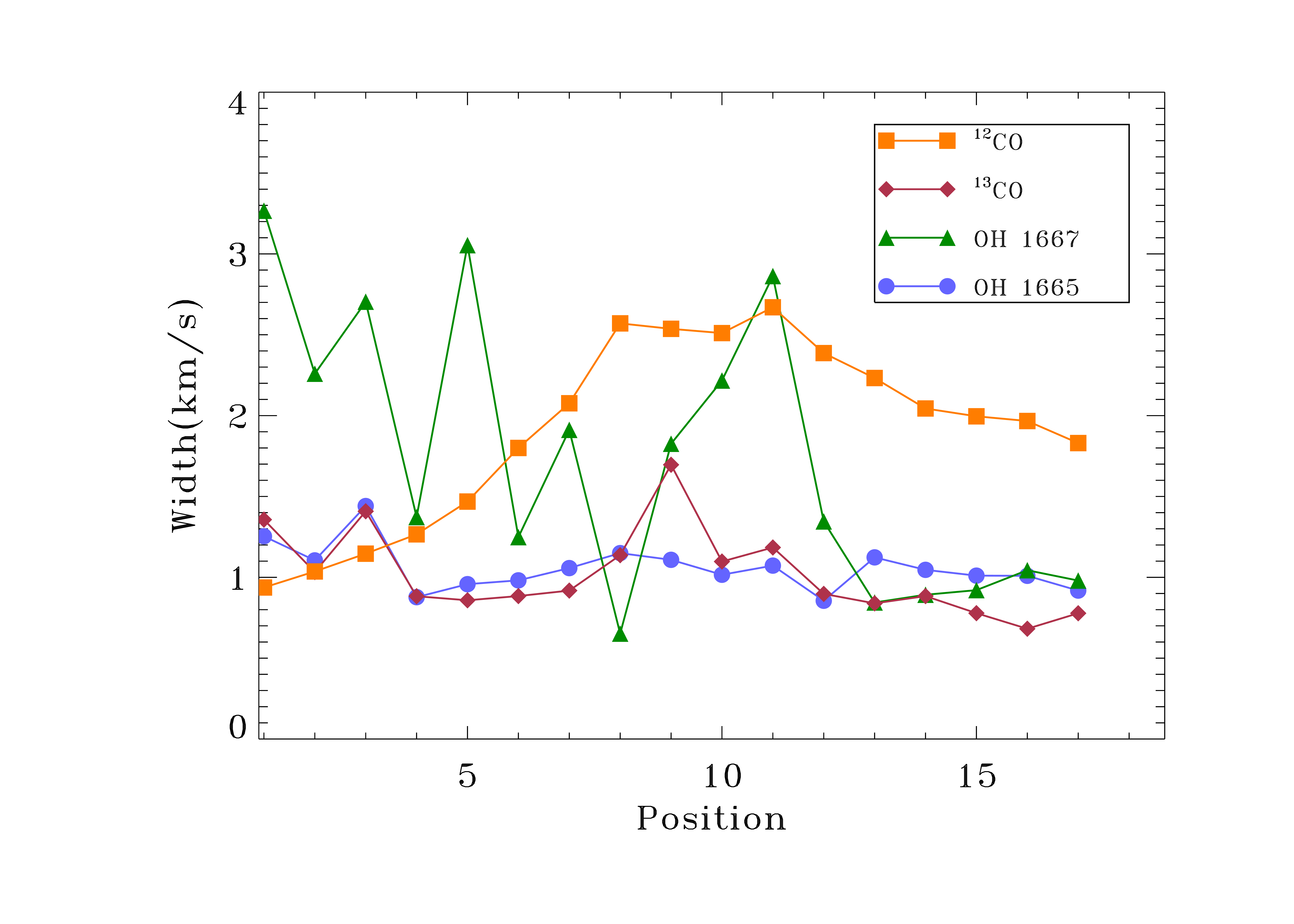}
\caption{The change of line width of OH 1665 MHz, 1667 MHz, \co~{J=1-0} and \13co~{J=1-0} along the cut direction shown in Fig.~1.}
\label{fig.line_width}
\end{figure}

We can also see the central velocities of component 1 in OH 1665 MHz shifting from 5.3 \kms~to 6.0 \kms~at positions 10-12 in Fig.~3. This velocity gradient may be caused by colliding streams or gas flow at the {TLB}. The description of the OH spectra and components across {TBR} can be found in Table~1.

\begin{table*}[htpb]
\centering

\caption{The change of OH spectral lines across the boundary in Taurus}
\label{tab_line_change1}
\begin{tabular}{cccccccccc}
\hline\\\vspace{-6.3mm}\\\hline\\
\multicolumn{1}{c}{Position} &\multicolumn{1}{c}{Offset} & \multicolumn{8}{c}{OH spectral lines} \\\cline{3-10}\\
 ID &   ( $'$ ) &\multicolumn{2}{c}{1612 MHz}    &  \multicolumn{2}{c}{1665 MHz} &  \multicolumn{2}{c}{1667 MHz} & \multicolumn{2}{c}{1720 MHz} \\
 
  & & $\mathfrak{C}$1$^{1}$ & $\mathfrak{C}$2$^{2}$ &$\mathfrak{C}$1 & $\mathfrak{C}$2 &$\mathfrak{C}$1 & $\mathfrak{C}$2 &$\mathfrak{C}$1 & $\mathfrak{C}$2 

  \\\hline   
1 (outer)                            & -24.0  &  $w\mathfrak{e}$$^{*3}$&  $\mathfrak{a}$$^{6}$  &$\mathfrak{n}$& $\mathfrak{e}$& $\mathfrak{n}$& $w\mathfrak{e}$& $\mathfrak{n}$ &$w\mathfrak{e}$    \\
7 ($\rm I_{H_2} peak$)    & -6.0    &  $w\mathfrak{e}$ &  $\mathfrak{a}$   & $\mathfrak{n}$&  $\mathfrak{e}$ & $w\mathfrak{e}$& $w\mathfrak{e}$ &$\mathfrak{n}$ & $\mathfrak{e}$  \\
9 ({boundary})                           &  0.0    &  $\mathfrak{e}$$^{4}$ & $w\mathfrak{a}$$^{7}$   & $\mathfrak{e}$&  $\mathfrak{e}$ & $\mathfrak{e}$& $w\mathfrak{e}$& $\mathfrak{a}$ & $\mathfrak{e}$ \\
11 ($\rm N_{H_2}$ peak) & 6.0    &  $\mathfrak{e}$($\rightarrow$$^{5}$) & $\mathfrak{n}$$^{8}$   & $\mathfrak{e}$($\rightarrow$) & $\mathfrak{e}$ & $\mathfrak{e}$($\rightarrow$)& $w\mathfrak{e}$& $\mathfrak{a}$($\rightarrow$) &$w\mathfrak{e}$ \\
12 ($\rm N_{CO}$ peak)  & 9.0    &   $\mathfrak{e}$($\rightarrow$) & $\mathfrak{n}$  &  $\mathfrak{e}$($\rightarrow$) & $w\mathfrak{e}$ &  $\mathfrak{e}$($\rightarrow$) & $\mathfrak{n}$&$\mathfrak{a}$($\rightarrow$) &$\mathfrak{n}$\\
17 (inner)                         &   24.0  &  $\mathfrak{e}$($\rightarrow$) & $\mathfrak{n}$  &  $\mathfrak{e}$($\rightarrow$)  &  $\mathfrak{n}$ &  $\mathfrak{e}$($\rightarrow$) & $\mathfrak{n}$&$\mathfrak{a}$($\rightarrow$) &$\mathfrak{n}$ \\\hline\\

\multicolumn{5}{p{.50\textwidth}}{ $^{1}$ $\mathfrak{C}$1 means the component at  5.3 \kms.} & \multicolumn{5}{p{.50\textwidth}}{ $^{5}$ $\rightarrow$ means the shifting of central velocity of component 1.}\\
\multicolumn{5}{p{.450\textwidth}}{ $^{2}$ $\mathfrak{C}$2 means the component at  6.8 \kms.}& \multicolumn{5}{p{.50\textwidth}}{ $^{6}$ $\mathfrak{a}$  means absorption. }\\
\multicolumn{5}{p{.450\textwidth}}{ $^{3}$ $w\mathfrak{e}$ means weak emission.}& \multicolumn{5}{p{.50\textwidth}}{ $^{7}$ $w\mathfrak{a}$  means weak absorption. }\\
\multicolumn{5}{p{.50\textwidth}}{ $^{4}$ $\mathfrak{e}$  means emission. }&\multicolumn{5}{p{.50\textwidth}}{ $^{8}$ $\mathfrak{n}$  means neither emission nor absorption. }\\

\end{tabular}
\end{table*}

\subsection{\13co Column Density Calculation}
\label{Column Density Calculation}

From the fitting result of \co~{J=1-0} and \13co~{J=1-0} above, we can calculate the column density of \13co. The column density of \13co\ in the upper-level ($J=1$) can be written
as
\begin{equation}
N_{u,^{13}\rm CO}=\frac{8\pi k\nu^2}{hc^3 A_{ul}}\int T_{b} (V) dV
\lc
\end{equation}
where $k$ is Boltzmann's constant, $h$ is Planck's constant, $c$ is the speed of light, $A_{ul}$
is the spontaneous decay rate from the upper level to the lower level, and $T_b$ is the brightness temperature.
A convenient form of this equation is
\begin{equation}
\left(\frac{N_{u,^{13}\rm CO}}{\rm cm^{-2}}\right)={3.6}\times
10^{14}\int \left(\frac{T_b}{\rm K}\right) {\rm d}\left(\frac{V}{\textrm{
km s$^{-1}$}}\right) \lp
\end{equation}
The total \13co\ column
density $N_{\rm tot}$ is related to the upper level column density
$N_{u}$ through (Li 2002)
\begin{equation}
N_{\rm tot, ^{13}CO}=f_{u} f_{\tau} f_b N_{u,\rm ^{13}CO} \lp
\end{equation}
In the equation above, the level correction factor $f_u$ can be
calculated analytically under the assumption of local thermal
equilibrium (LTE) as
\begin{equation}
f_{u}=\frac{Q(T_{\rm ex})}{g_u \exp\left(-\frac{h\nu}{kT_{\rm ex}}\right)} \lc
\end{equation}
where $g_u$ is the statistical weight of the upper-level. $T_{\rm
ex}$ is the excitation temperature and $Q(T_{\rm ex})=kT_{\rm ex}/hB_0$ is the LTE
partition function, where $B_0$ is the rotational constant
(Tennyson 2005). A convenient form of
the LTE partition function is $Q(T_{\rm ex})\approx T_{\rm ex}/2.76\rm K$.
The correction factor for opacity is defined as
\begin{equation}
f_{\tau}=\frac{\int\tau_{13}dv}{\int(1-e^{-\tau_{13}}){\rm d}v} \lc
\end{equation}
and the correction for the background
\begin{equation}
f_b=\left[1-\frac{e^{\frac{h\nu}{kT_{\rm
ex}}}-1}{e^{\frac{h\nu}{kT_{\rm bg}}}-1}\right]^{-1} \lc
\end{equation}
where $\tau_{13}$ is the opacity of the $^{13}$CO transition and
$T_{\rm bg}$ is the background temperature, assumed to be 2.7K.

The \13co\ opacity is estimated as follows. Assuming equal excitation temperatures for the two
isotopologues, the ratio of the brightness temperature of $^{12}$CO to that of $^{13}$CO
can be written as
\begin{equation}
\frac{T_{b, 12}}{T_{b, 13}}=\frac{1-e^{-\tau_{12}}}{1-e^{-\tau_{13}}} \lp
\end{equation}
Assuming $\tau_{12}\gg 1$, the opacity of $^{13}$CO can be written as
\begin{equation}
\tau_{13}=-\ln\left(1-\frac{T_{b,13}}{T_{b,12}}\right) \lp
\end{equation}

The excitation temperature $T_{\rm ex}$ is obtained from the $^{12}$CO
intensity. First, the maximum intensity in the spectrum of each
pixel is found. This
quantity is denoted by $T_{\rm max}$. The excitation
temperature is calculated by solving the following equation
\begin{equation}
T_{\rm max}=\frac{h\nu}{k}\left[\frac{1}{e^{\frac{h\nu}{kT_{\rm
ex}}}-1}-\frac{1}{e^{\frac{h\nu}{kT_{\rm bg}}}-1}\right] \lc
\end{equation}
where $h$, $k$ and $\nu$ are Planck's constant, Boltzmann's
constant, and the central frequency of $^{12}$CO $J=1\to 0$ line
(115.27 GHz), respectively.

{To examine the LTE assumption when calculating the physical parameters of CO, we used a spherical 1-d non-LTE spectral analysis radiative transfer model, RADEX~{(van der Tak et al. 2007)} to derive the excitation temperature of \co\ and \13co. We took position 10 as an example. We assumed the number density of \h2\ to be 400 \cm3, which is given by Orr et al. (2014). We also assumed the kinetic temperature to be 15 K, which is widely applied in the relatively diffuse region in the Taurus molecular cloud (e.g. Goldsmith et al. 2008). The resulting excitation temperature of \co\ and \13co\ are T$_{\rm ex,^{12}co}$ = 7.7 K and T$_{\rm ex,^{13}co}$ = 6.8 K, with a difference of about 10\%. The assumption that the excitation temperatures of \co\ and \13co are equal seems reasonable. Owing to the difference of excitation temperature between \co\ and \13co, the derived column density of \13co also has an error about 10\%.}

We show the change of excitation temperature of \13co and the change of column density of \13co along the cut direction in Table~2 and Fig.~6. The excitation temperature $T_{\rm ex}$ of \13co\ increases crossing the boundary. The column density of \13co\ shows a peak at position 13 and 14 inside the boundary.

\begin{figure}[htp]
\includegraphics[width=1.0\linewidth]{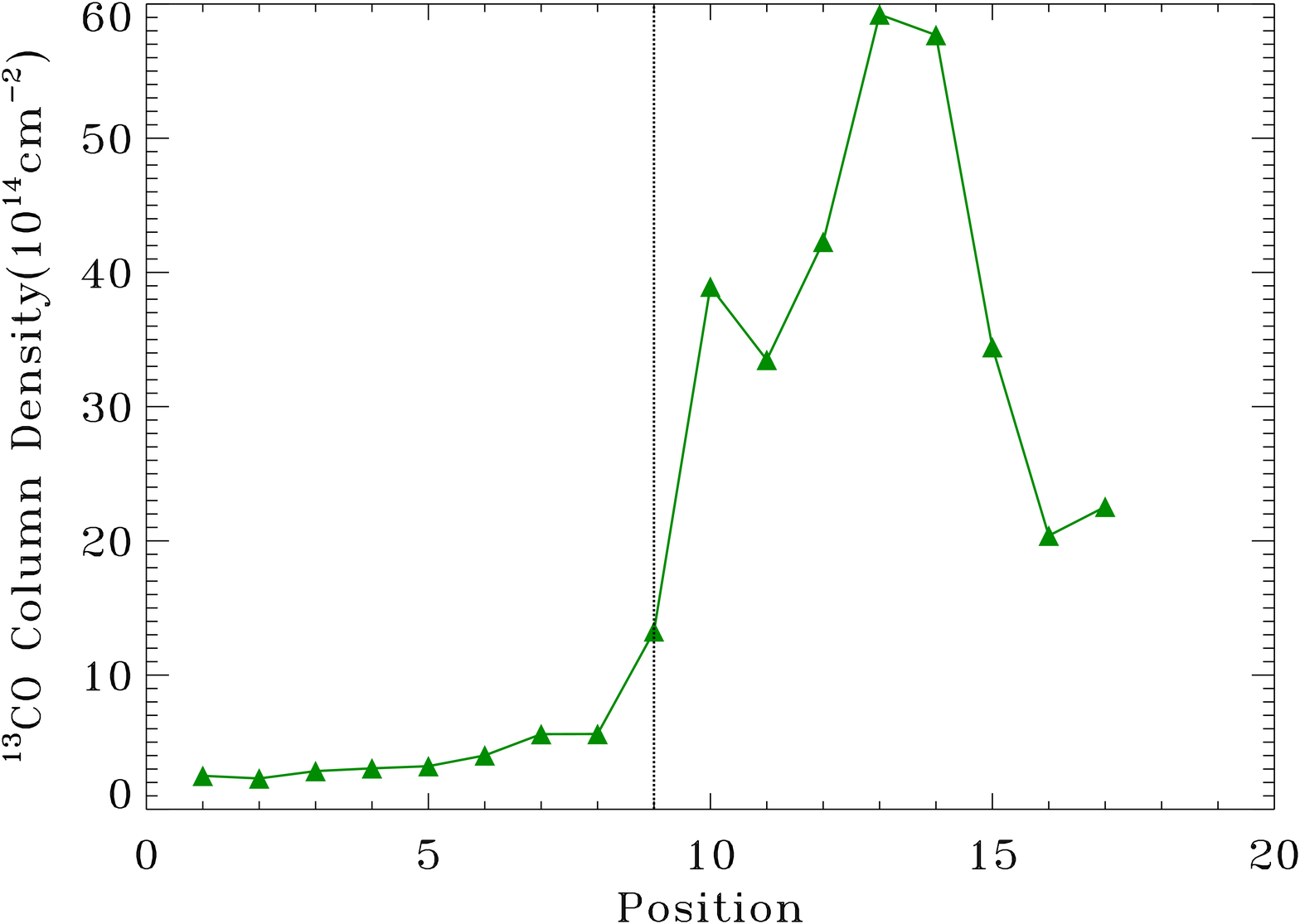}
\caption{The change of column density of \13co~along the cut direction.}
\label{fig.colden}
\end{figure}

\begin{table}[htb]
\begin{center}
\caption{parameters along the boundary  \label{tab.parameter}}
\begin{tabular}{cccc}
\hline
    position  & T$_{\rm ex,^{13}co}$   & N$_{^{\rm 13}\rm co}$  \\
      & (K)   & ($10^{14}\rm cm^{-2}$) \\
\hline
    1     & 6.5   & 2.5 \\
    2     & 6.3   & 2.3 \\
    3     & 5.9   & 2.8 \\
    4     & 5.8   & 3.1 \\
    5     & 5.6   & 3.2 \\
    6     & 5.6   & 4.0 \\
    7     & 6.3   & 5.6 \\
    8     & 6.3   & 5.6 \\
    9     & 6.9   & 13 \\
    10    & 7.6   & 39 \\
    11    & 7.8   & 33 \\
    12    & 7.7   & 42 \\
    13    & 7.6   & 59 \\
    14    & 8.5   & 58 \\
    15    & 8.5   & 34 \\
    16    & 8.0   & 20 \\
    17    & 8.5   & 23 \\
\hline
\end{tabular}
\end{center}
\end{table}

 \section{Simulation of OH Emission Line with RADEX}
 \label{Simulation of OH Emission line with RADEX}
We assume a cylindrical geometry to model the linear {boundary} region of the cloud. We adopted the results of ``cylindricalized" Meudon PDR model by Orr et al. (2014) and applied a spherical 1-d non-LTE spectra analysis radiative transfer model, RADEX~{(van der Tak et al. 2007)}, to generate a sky-plane image of the {TBR} for the transitions of OH.

\subsection{Density Profile of Taurus {Boundary} Region}
\label{Linear Edge Density Profile}
We modeled the linear {boundary} as a cylinder with a radius-dependent H$_2$ volume density structure of the following form (King 1962),
\begin{equation}
   n_{\rm H_2}(r) = \left\{
     \begin{array}{lr}
       n_ca^2/(r^2 + a^2) & : r \leq R\\
       0 & : r > R 
     \end{array} .
   \right.
\label{eq:king}
\end{equation}

This functional form was used in analysis of the Taurus region by Pineda et al. (2010).

This profile has a flat high-density center, which then transitions to a region of power-law decay, and finally, at a truncating radius, the density is set to zero.  It assumes only three parameters, the central core density $n_c$, a parameter characterizing the width of the central core $a$, and the truncating radius $R$. Orr et al. (2014) fitted the profile with visual extinction data for this region. The fitted values for the density profile parameters are $n_c = 626$ cm$^{-3}$, $a =0.457$ pc, and $R = 1.80$ pc.  The axis of the cylindrical distribution is centered $11.5'$ to the southwest of the {boundary} position between position 12 and position 13.

The \h2 volume density of 17 observed points along the boundary is obtained from Orr et al. (2014).

\subsection{{RADEX Fitting Results} }
\label{Simulation of OH Emission line with RADEX sub}

RADEX takes the following inputs: kinetic temperature ($T_{\rm k}$), density of \h2 ($n_{\rm H_2}$), \h2~ortho-to-para ratio (OPR), background temperature ($T_{\rm bg}$), column density of OH ($N_{\rm OH}$), and line width ($\Delta \mathbf v_{\rm OH}$). {Rates for collisional excitation of OH are taken from Offer et al. (1994). }

The $n_{\rm H_2}$ is an input parameter estimated based on the 
cylindrical model in Orr et al. (2014). The Galactic background emission is estimated to be about 0.8 K by extrapolating the standard interstellar radiation field (ISRF) to L band (Winnberg et al. 1980). $T_{\rm bg}$=3.5 K is thus used in the simulation. 

We vary the $T_{\rm k}$, \h2~OPR and $N_{\rm OH}$ to find the optimum model by minimizing $\chi^2$ for the four OH lines, defined as
\begin{equation}
\chi^2=\frac{1}{N}\sum\limits_{i=1}^N \frac{(I_{model_i}-I_{obs_i})^2}{\sigma_{obs_i}^2},
\end{equation}
where $I_{obs}$ are the four observed OH lines' instensities, $I_{model}$ are the model line generated by RADEX, $\sigma_{obs}^2$ are the RMS of the four observed OH lines.

We varied $T_{\rm k}$, OPR and $N_{\rm OH}$ to obtain the best fit to the observation at position 1, shown in Fig.~7. The best fitting $T_{\rm k}$, OPR and $N_{\rm OH}$ are 31 K, 0.2 and $3.7\times 10^{14}\; \rm cm^{-2} $, respectively. We also calculated the column density of OH at position 1 (assuming it to be optical thin and with no background emission) from the integrated intensity
(in K \kms) of the 1667 MHz line through (e.g. Knapp \& Kerr 1973; Turner \& Heiles 1971)
\begin{equation}
\label{noh}
\frac{N_{\rm {{OH}}}}{\textrm{cm}^{-2}} = 2.22\times10^{14}\frac{\int {T_{b} (V) dV}}{\textrm{K km s$^{-1}$}} \lp
\end{equation}	
With these assumptions, $N_{\rm OH}$ is $0.5\times 10^{14}$ \cms. When the excitation temperature of OH is so low that the background cannot be neglected, we have a correction factor $f_{bg}$ (e.g. Harju et al. 2000; Suutarinen et al. 2011)
\begin{equation}
\label{fbg}
f_{bg}=\frac{1}{1-T_{\rm bg}/T_{\rm ex}} \lp
\end{equation}	
When $T_{\rm bg}$ $\simeq$ 3.5 K and $T_{\rm ex}$ $\simeq$ 4 K, $f_{bg}$ $\simeq$ 8, which yields $N_{\rm OH}$ $\simeq$ $4\times 10^{14}~\rm {cm^{-2}}$, almost the same as the fitted $N_{\rm OH}$ from RADEX (denoted $N_{\rm OH}^{RADEX}$). The problem with using the ``{LTE} method'' is that we do not know the excitation temperature of OH. Instead, we have to assume an excitation temperature for OH, which has a major effect on the correction factor $f_{bg}$. Only with a statistical equilibrium calculation (e.g. RADEX) can we know the excitation temperature of OH exactly. {We should be cautious about the low excitation temperature of OH, otherwise we may underestimate the column density of OH by a factor of 8 or more in many conditions where the LTE method is used.} 

Compared with $N_{\rm OH}^{RADEX}$, the calculated $N_{\rm OH}$ through Eq.~12 and Eq.~13 {(noted as $N_{\rm OH}^{LTE}$ representing the ``LTE method")} is almost the same as $N_{\rm OH}^{RADEX}$. The fitted opacity (Table~3) based on the model is relatively smaller ($\tau\sim$ 0.04-0.3), which is consistent with the assumption of optical thin in the ``{LTE} method". 

\begin{figure*}[htpb]
\includegraphics[width=1.0\linewidth]{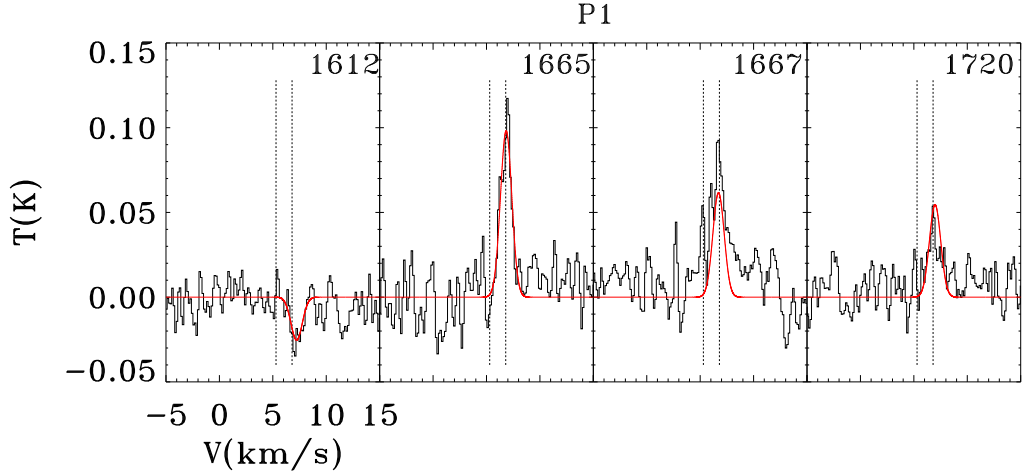}
\caption{Observed OH lines(black lines) and the simulated OH profiles(red lines) using RADEX at position 1.}
\label{fig.add_oh_radex1}
\end{figure*}

\begin{table}[htbp]
  \centering
  \caption{OH physical parameters at position 1
    \label{tab.oh1}}     
    \begin{tabular}{cccc}
    \hline
   \multicolumn{4}{c}{{n$_{\rm H_{2}}$}=62.6 \; \rm \cm3  \;\; \rm OPR=0.20\; \; $T_{\rm k}=31.0$ K }\\
   \multicolumn{4}{c}{$N_{\rm OH}=3.7\times 10^{14}\; \rm cm^{-2} $}\\  
    \hline
    line  &~~~$T_{\rm ex}$ & $\tau$  &   $T_{\rm A} $ \\
      & (K)   & & (K) \\
\hline
1612 &  3.1  &  0.045& -0.025     \\
1665 & 4.1  &  0.163& 0.098         \\
1667 &  3.8  &  0.331& 0.062            \\
1720 &  5.4  &  0.024&0.055           \\
\hline
    \end{tabular}%
\end{table}%

Eq.~12 used to calculate $N_{\rm OH}$ assumes LTE, but the observed line ratio between OH 1665 MHz and 1667 MHz is obviously greater than 1, indicating that the OH at position 1 is far from LTE. {The critical density of OH {HFS lines} at 10-50 K is 1-20 \cm3, indicating that OH excitation is dominated by collision in TBR with n=70-600~\cm3. Compared with OH, {low-J \co\ lines have} a larger critical density about 2000 \cm3. In most cases, when collision dominate the excitation, the LTE assumption is reasonable. Surprisingly, OH {HFS lines are} far from LTE, which has a low critical density. \co\ is consistent with LTE as discussed in Section~3.2 with RADEX analysis. Considering for the complex energy level of OH, some pumping mechanism must be operative to yield the non-LTE of OH. We will have a detailed discussion in Section~5. A most possible mechanism is C-shock.} The line ratio between 1665 MHz and 1667 MHz, and the integrated intensity of 1667 MHz across the boundary are listed in Table~4. In {TBR-O}, the line ratio is greater than 1. In {TBR-I}, the line ratio {is} between $\simeq$ 0.5 and $\simeq$ 0.8, within the range allowed by LTE.

\begin{table}[htb]
\begin{center}
\caption{OH parameters along the boundary  \label{tab.oh}}
\begin{tabular}{ccc}
\hline
    position  & Ratio(1665/1667)   & Intensity(1667)  \\
      &    & ($\rm K~km~s^{-1}$) \\
\hline
    1     & 1.6 $\pm$~0.18   & 0.21 $\pm$~0.02 \\
    2     & 1.2 $\pm$~0.14  & 0.20 $\pm$~0.02\\
    3     & 1.3 $\pm$~0.13   & 0.22 $\pm$~0.02 \\
    4     & 1.2 $\pm$~ 0.12  & 0.24 $\pm$~0.02 \\
    5     & 1.5 $\pm$~0.14   & 0.29 $\pm$~0.02 \\
    6     & 1.6 $\pm$~0.13   & 0.26 $\pm$~0.02 \\
    7     & 1.3 $\pm$~0.10   & 0.35 $\pm$~0.02 \\
    8     & 1.2 $\pm$~0.08  & 0.31 $\pm$~0.02 \\
    9     & 1.1 $\pm$~0.05  & 0.42 $\pm$~0.02 \\
    10    & 0.51 $\pm$~0.02  & 0.49 $\pm$~0.01 \\
    11    & 0.54 $\pm$~0.02  & 0.53 $\pm$~0.02 \\
    12    & 0.52 $\pm$~0.02  & 0.50 $\pm$~0.01 \\
    13    & 0.61 $\pm$~0.02  & 0.49 $\pm$~0.01 \\
    14    & 0.69 $\pm$~0.02  & 0.48 $\pm$~0.01 \\
    15    & 0.72 $\pm$~0.02  & 0.42 $\pm$~0.01 \\
    16    & 0.77 $\pm$~0.03  & 0.40 $\pm$~0.01 \\
    17    & 0.83 $\pm$~0.04  & 0.37 $\pm$~0.01 \\
\hline
\end{tabular}
\end{center}
\end{table}

The change of physical parameters across the boundary is partially listed in Table~5. 

\begin{table*}[htpb]
\centering

\caption{The change of physical parameters across the boundary in Taurus}
\label{tab_line_change2}
\begin{tabular}{cccccccccc}
\hline\\\vspace{-6.3mm}\\\hline\\
Position &\multicolumn{1}{c}{Offset} & \multicolumn{7}{c}{Physical Parameters } \\\cline{3-9}\\
 ID &   ( $'$ ) & \multicolumn{2}{c}{{$\rm T_k$} (K) }   & \multicolumn{2}{c}{ {$\rm N_{OH}$}} ($10^{14}$ cm$^{-2}$) & {A$_v$ }(mag) & $\rm {N_{OH}/N_{H_2}}$ $(10^{-7})$ & DGF$^{1}$ \\
 
  & &$\mathfrak{C}$1 & $\mathfrak{C}$2 &$\mathfrak{C}$1 & $\mathfrak{C}$2 & & & &

  \\\hline   
1 (outer)                            & -24.0 &  -- & 37 & --& 2.9   & 0.4   & 7.2$^{+3}_{-2}$& 0.75$^{+0.1}_{-0.1}$\\
7 ($\rm I_{H_2} peak$)    & -6.0     &25&35  & 2.9&1.3   & 1.3   & 3.5$^{+0.7}_{-0.3}$& 0.70$^{+0.1}_{-0.1}$\\
9 ({boundary})                           &  0.0    &26&32  & 1.5&4.9   & 1.8   & 3.7$^{+0.4}_{-0.2}$& 0.13$^{+0.2}_{-0.3}$\\
11 ($\rm N_{H_2}$ peak) & 6.0    &10&27  & 1.5&4.5   & 2.4   & 2.6$^{+0.5}_{-0.3}$ & 0.03$^{+0.2}_{-0.4}$\\
12 ($\rm N_{CO}$ peak)  & 9.0   &23&26  & 2.1&2.5   & 2.6   & 1.9$^{+0.4}_{-0.2}$  & 0.07$^{+0.2}_{-0.4}$\\
17 (inner)                         &   24.0  &24&--  & 3.3&--   & 1.5   & 2.3$^{+0.5}_{-0.3}$ & 0.12$^{+0.2}_{-0.3}$\\\hline\\
\multicolumn{7}{p{.450\textwidth}}{ $^{1}$ DGF means dark gas fraction.}\\
\end{tabular}
\end{table*}

\subsection{The Effect of the \h2~Ortho-to-Para Ratio (OPR) on Fitting}
\label{The Effect of OPR in Fitting}

Owing to the different cross sections between OH and two spin {symmetries} of \h2, i.e. ortho-\h2 and para-\h2\ {(Offer et al. 1994)}, {the} OPR plays a significant role in the excitation of OH. The exact value of OPR is important for producing the observed OH 1665/1667 intensity ratio in certain positions, such as for position 9, as shown in Fig.~8. Overall, the derived column density is insensitive to the numerical value of OPR, such as for position 10, as shown in Fig.~8. We will discuss this issue in a separate paper and focus on the dark gas and OH abundance content in the present work.

\begin{figure}[htp]
\includegraphics[width=0.98\linewidth]{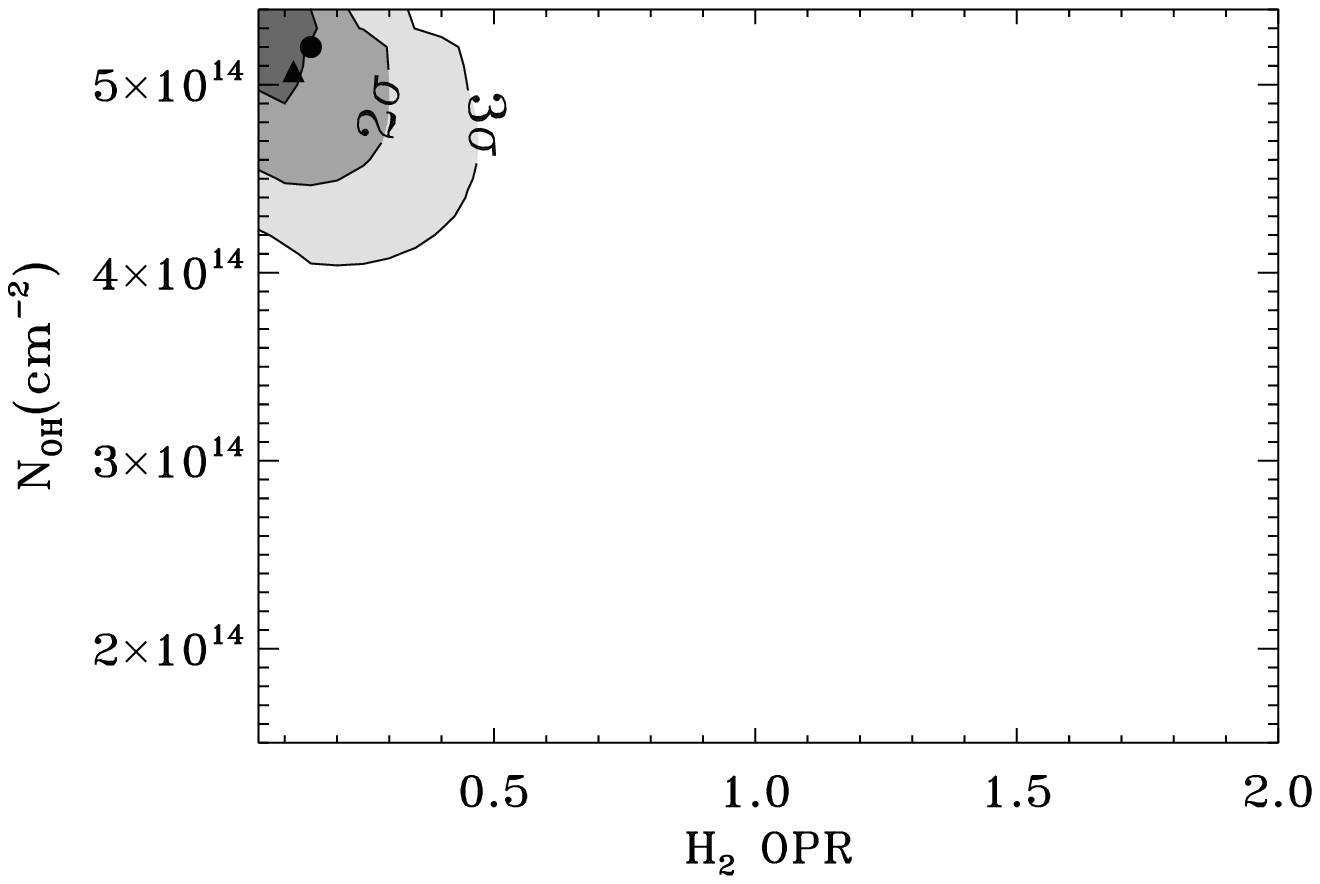}
\includegraphics[width=0.98\linewidth]{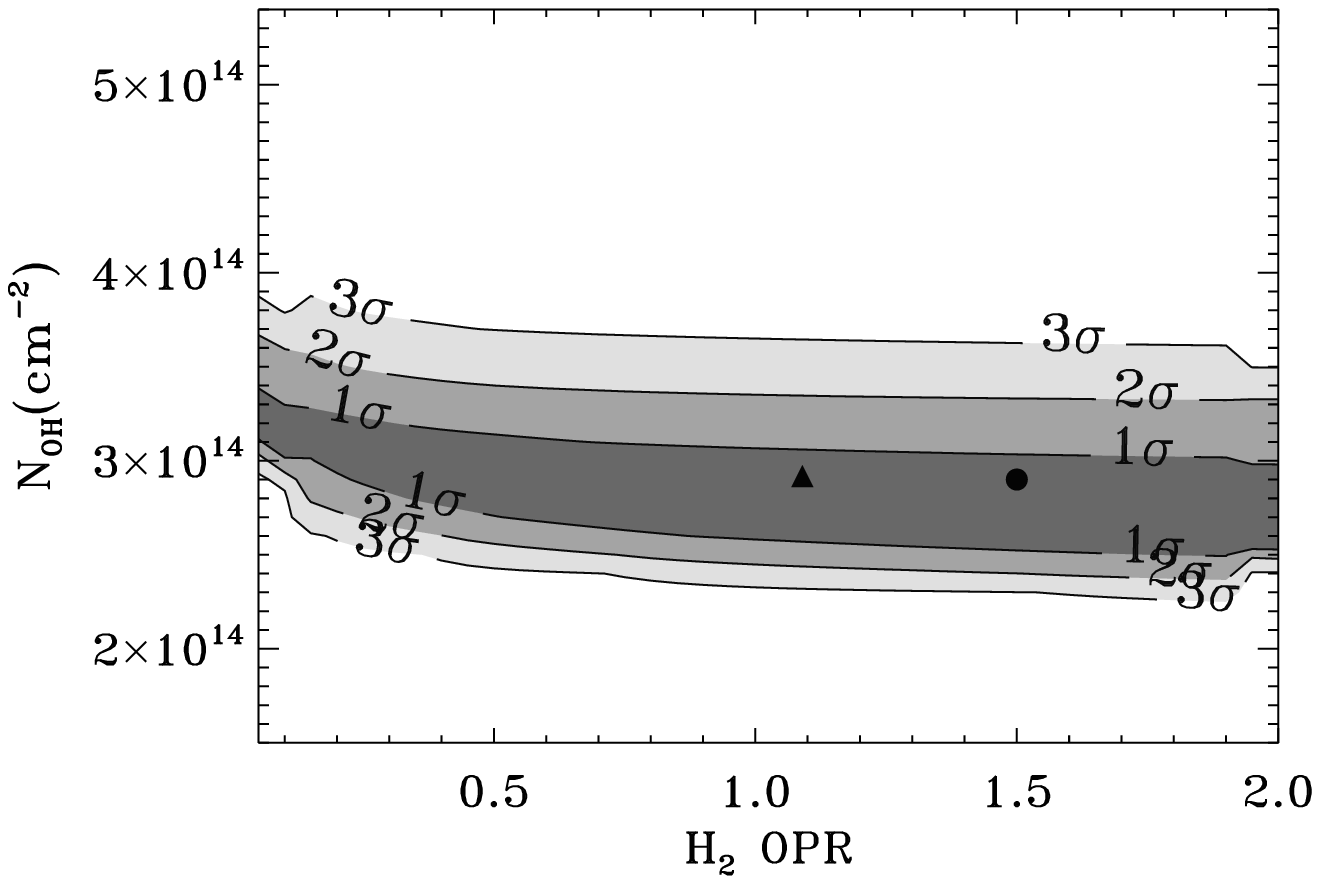}
\caption{The probability space of each parameter at position 9 (top panel) and position 12 (bottom panel). In the contour map of probability space of two different parameters, different shades represent different confidence levels. The dot represents the $\chi ^{2}_{min}$ fitted parameter, and the triangle represents the mean value of the parameter by integrating the parameter in probability space.}
\label{fig.para_9}
\end{figure}

\subsection{{Physical Parameter Analysis}}
\label{Physical Parameters analysis}

We define the {CO-dark molecular}~gas fraction or dark gas fraction (DGF) as

\begin{equation}
{\rm {DGF}} =1-\frac{N_{\rm CO}\times 10^4}{N_{\rm H_2}} \lc
\end{equation}
which represents the fraction of \h2 that cannot be traced by CO emission. $N_{\rm CO}$ is obtained from the LTE calculation in Section~3.2 with the assumption that the abundance ratio of \co\ to \13co\ is 65 (Langer \& Penzias 1993; Liszt 2007). $N_{\rm H_2}$ is obtained from integration of the density profile in Section~4.1 alone line of sight and the visual extinction is obtained from the relation $N($H$_2)/A_{\rm V}=9.4\times10^{20}$ cm$^{-2}$ mag$^{-1}$, by assuming that the hydrogen is predominately in molecular form {and standard grain properties are appropriate for the diffuse ISM} (Orr et al. 2014). We averaged the column density of \h2\ within the OH beam size ($\sim$ 3$^{\prime}$) at each position to make all the calculations refer to the same beam size. The variation of  $T_{\rm k}$,  $N_{\rm OH}$, $N_{\rm OH}/N_{\rm H_2}$ and DGF across the boundary {as a function} of extinction A$_{v}$ is shown in Fig.~9. When the extinction A$_{v}$ increases from 0.4 to 2.6 magnitudes, the kinetic temperature, the abundance of OH and the dark gas fraction all decrease. Especially, {the DGF} decreases from 80\% to 20\%, which means the amount of  molecular gas that cannot be traced by CO is three times larger than that of molecular gas traced by CO when the extinction is below 1.4 magnitude. Empirically, CO intensities have been used as an indicator of the total molecular mass in the Milky way and in galaxies through the so-called ``X-factor" with numerous well-known caveats. In other words, we may seriously underestimate the amount of molecular gas through ``X-factor" in low extinction clouds or regions of galaxies. 

\begin{figure}[htp]
\includegraphics[width=1.0\linewidth]{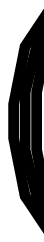}
\caption{The change of $T_{\rm k}$, $N_{\rm OH}$, $N_{\rm OH}/N_{\rm H_2}$ and DGF across the boundary as a function of the extinction A$_{v}$. }
\label{fig.mp_av_all}
\end{figure}

We parametrize the trend of $N_{\rm OH}/N_{\rm H_2}$ and DGF in an exponential law and a gaussian profile, respectively, as shown in Fig.~10. The trend of $N_{\rm OH}/N_{\rm H_2}$ can be fitted as
 \begin{equation}
 \label{ohh2_eq}
\frac{N_{\rm OH}}{N_{\rm H_2} }=1.5\times10^{-7}+0.9\times10^{-7}\times \exp(-\frac{A_{v}}{0.81}) .
\end{equation}
When the visual extinction is much larger than 0.48 mag, $N_{\rm OH}/N_{\rm H_2}$ remains roughly a constant $1.5\times10^{-7}$. This constant indicates the abundance of OH at large visual extinctions within molecular clouds. A calculation using the Meudon PDR Code (Le Petit et al. 2006) with conditions appropriate for the {TBR} yields reasonably good agreement with OH abundance [OH]/[\h2] for moderate extinction (A$_v$~$\sim$~2).  The prediction of [OH]/[\h2] at extinctions at or below 1 mag is far lower {(by a factor of 80)} that we derive, suggesting that there may be an additional channel of OH production active, possibly due to the shock (e.g. Draine \& Katz 1986) produced by the colliding streams. {When shock waves propagate through the molecular ISM the ambient gas is compressed, heated, and accelerated. When temperature is above 􏰂300 K, the neutral-neutral reactions become important, which yield the overabundance of OH (Neufeld et al. 2002).}


The trend of DGF can be fitted as,
 \begin{equation}
 \label{DGF_eq}
{\rm DGF}=0.90\times \exp[-(\frac{A_{v}-0.79}{0.71})^2].
\end{equation}
The peak of DGF is located at A$_{v}$=0.79 mag, and the FWHM is 1.4 mag. When 0.79 $<$A$_{v}<2.5$ mag, the DGF decreases sharply with increasing A$_{v}$. This behavior of the DGF is expected theoretically, {e.g.~Planck and Fermi Collaborations (2015) found that {CO-dark molecular} dominates the molecular columns at A$_v$~$\le$~1.5 mag and Paradis et al. (2012) found an obvious deviation of the amount of gas traced by \hi\ and \co\ from that of the extinction data in the extinction range between 0.3 and 2 mag, while neither the low extinction ($\le$~0.3 mag) nor the high extinction ($\ge$~2 mag) showing significantly deviation.} In regions with sufficiently low visual extinction, \h2 cannot survive due to the destruction by UV photons. On the other hand, when the visual extinction is high enough, \co ~can be formed and survive. At this point, the molecular gas can be well traced by \co, so the DGF also drops at high visual extinction. When the visual extinction is between 0.4 and 1.3 mag, the abundance of \h2 is already a layer fraction of the total hydrogen abundance but \co ~has not formed completely, so the DGF peaks under these conditions. 

It is important to emphasize that the parameterized trends above can only be applied to the molecular gas with visual extinction between 0.4 and 1.3 mag, for which we have OH data and have modeled the lines successfully.

\begin{figure}[htp]
\includegraphics[width=1.0\linewidth]{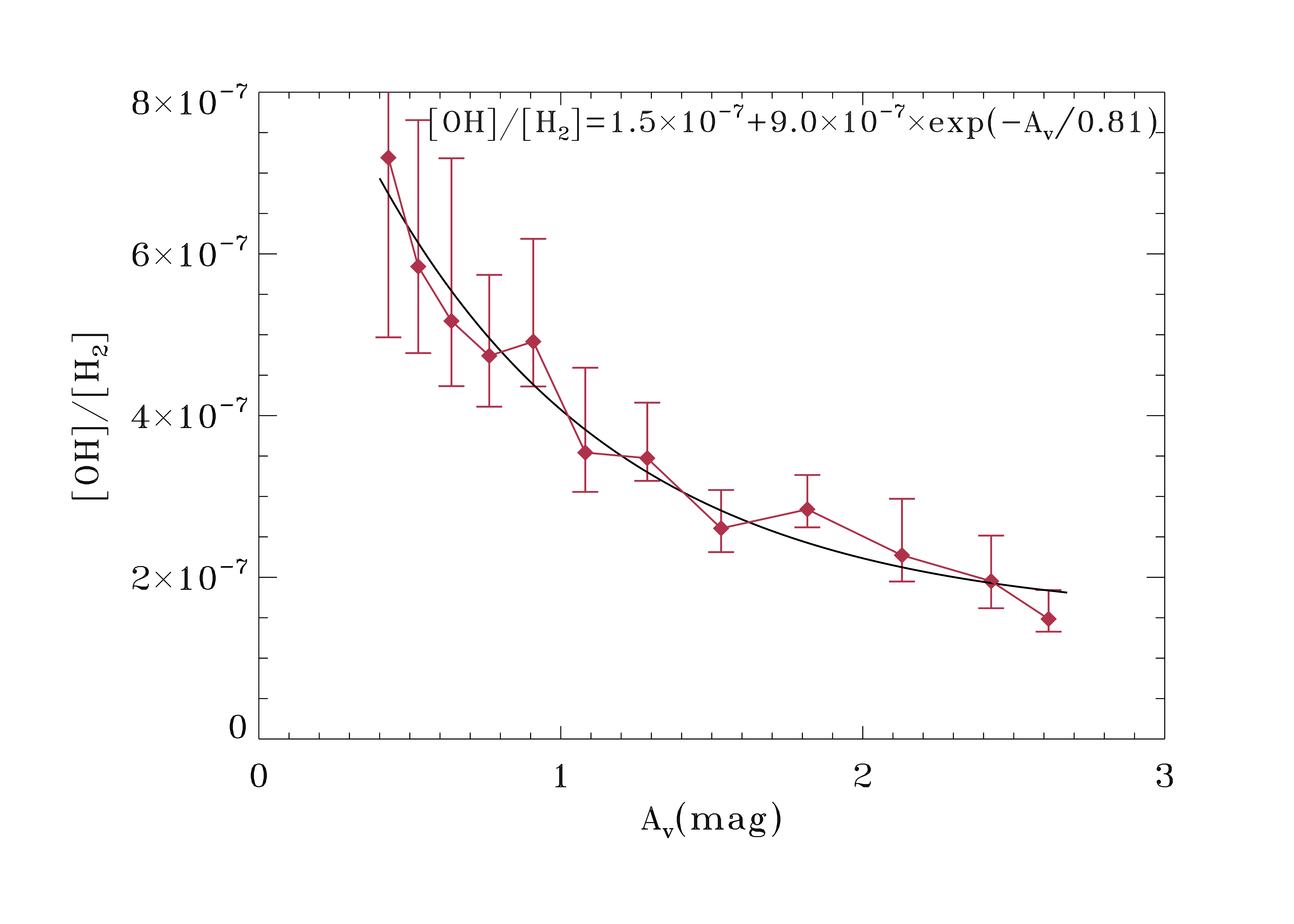}
\includegraphics[width=1.0\linewidth]{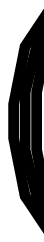}
\caption{Parameterization of the change of $N_{\rm OH}/N_{\rm H_2}$ and DGF across the boundary as a function of extinction A$_{v}$. In the bottom panel of DGF plot, the value of DGF becomes slightly negative when A$_{v} \ge 1.8$~mag (the last three points) owing to there being slightly more \h2 traced by CO than that calculated from extinction. As the error in A$_{v}$ increases at high extinction (by almost 50\%, as shown in Fig.~2 in Orr et al.\ (2014)), we rescale $N_{\rm H_2}$ when A$_{v} \ge 1.8$~mag (the last three points) to be close to zero by the same factor of 1.6 in the bottom panel to ensure that DGF $\ge$ 0. }
\label{fig.fit_oh_dg}
\end{figure}

\section{Conjugate Emission of OH}
\label{Conjugate Emission of OH}

As shown in Fig.~2, OH 1612 MHz lines appear to be absorption lines and OH 1720 MHz lines appear to be emission lines outside the boundary. However, it is quite different inside the boundary. OH 1612 MHz lines appear to be emission lines and OH 1720 MHz lines appear to be absorption lines inside the boundary. This inversion of satellite lines is known as conjugate emission of OH.

The mechanism for producing conjugate emission is asymmetry in pumping due to quantum selection rules in the OH rotational transition
ladder, shown in Fig.~11. 
The 18~cm OH maser lines result from hyperfine transitions in the 
$^2\Pi_{3/2}(J=3/2)$ level, which is the ground state. The 1720~MHz line 
is produced by a transition from a $F=2$ to $F=1$ state, while 
the 1612~MHz line is a transition from $F=1$ to $F=2$ .
Transitions between rotational levels are
permitted when $|\Delta F| = 0, 1$. The $^2\Pi_{3/2}(J=5/2)${-$^2\Pi_{3/2}(J=3/2)$ intra-ladder transition at $\lambda$~=~119~$\mu m$ } has hyperfine
levels with $F={2, 3}$, and thus will preferentially populate the $F=2$ levels in
the ground state. The result is 1720~MHz inversion, and anti-inversion of 
the 1612~MHz line. A radiative transition from $^2\Pi_{1/2}(J=1/2)$ {-$^2\Pi_{3/2}(J=3/2)$ cross-ladder transition at $\lambda$~=~79~$\mu m$}, which has hyperfine levels with $F={0,1}$, will overpopulate 
the $F=1$ levels in the OH ground state, producing 1612~MHz inversion and 
1720~MHz anti-inversion. 

\begin{figure}[htpb]
    \centering
    \includegraphics[width=3.5in] {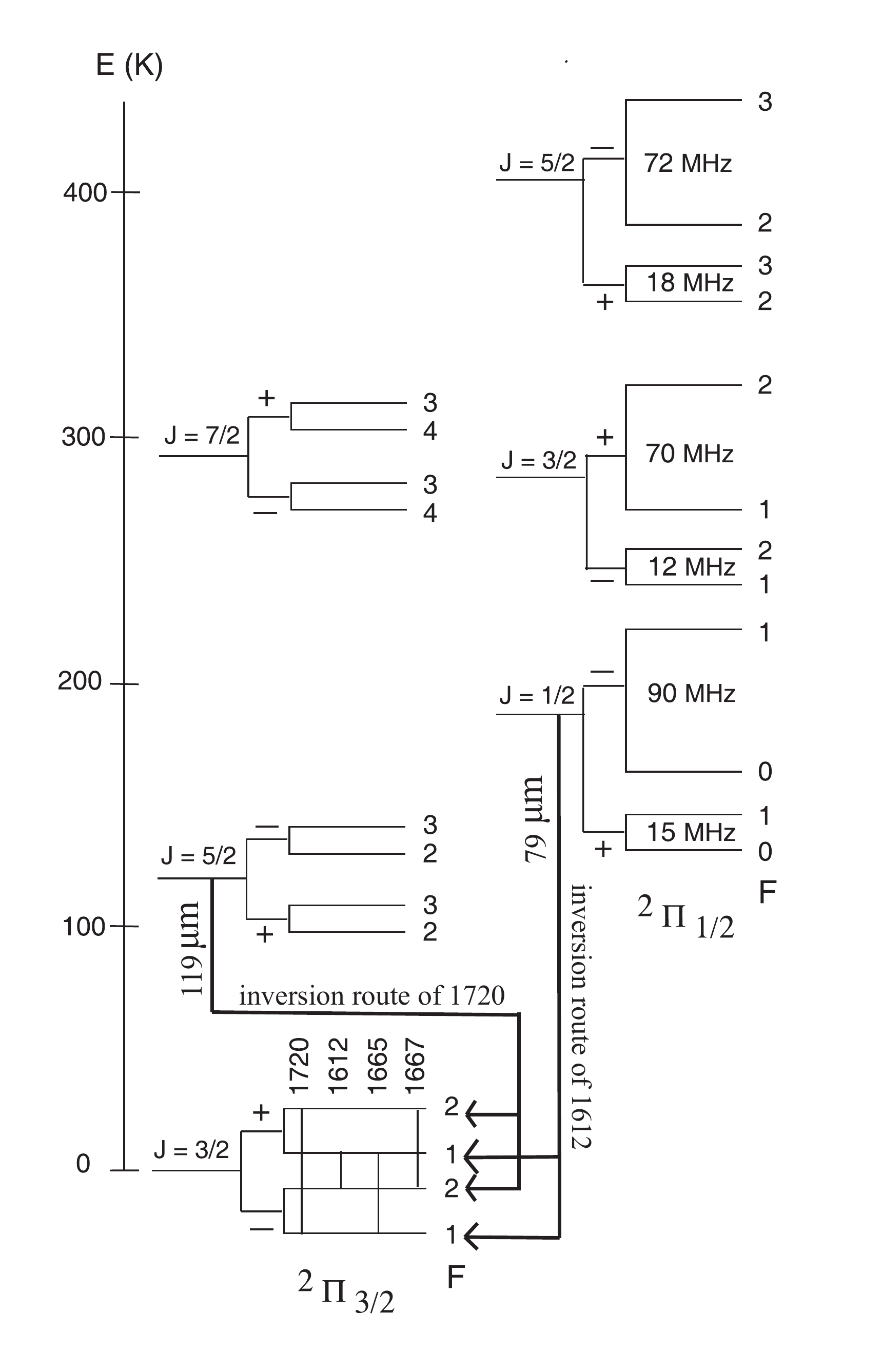}
    \caption{Rotational energy levels of OH, based on Lockett \& Elitzur (2008). There are four transitions within each rotational level as a result of $\Lambda$-doubling and hyperfine splitting (not shown to scale).} 
\label{fig:energy_level}
\end{figure}

\begin{figure*}[htpb]
\includegraphics[width=1.0\linewidth]{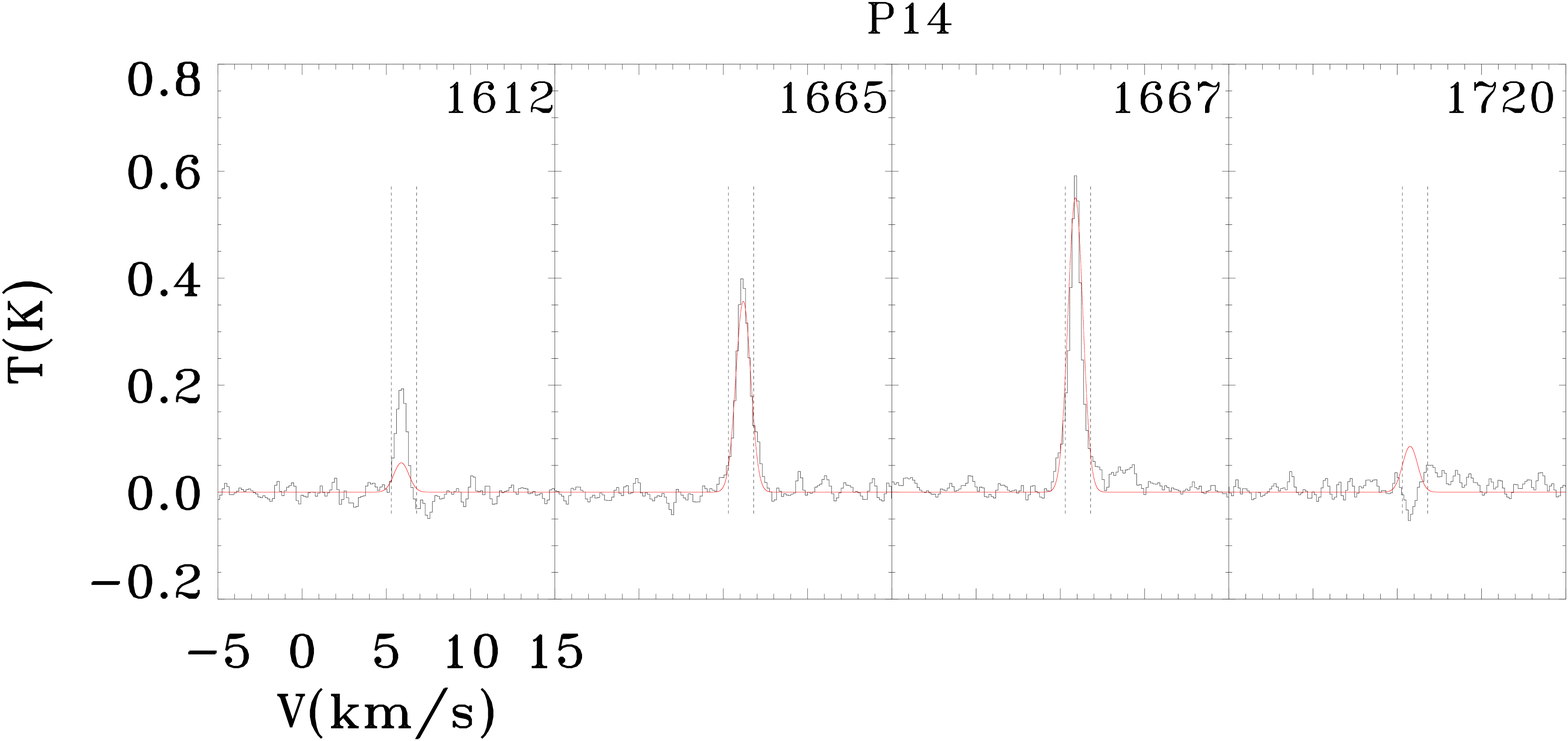}
\caption{The panel shows the observed OH lines(black lines) and the simulated OH profiles(red lines) with RADEX at position 14.}
\label{fig.add_oh_radex14}
\vspace{5mm}
\end{figure*}

For either of these pumping mechanisms, the FIR
transition must be optically thick. When both FIR lines are optically thick,
the 1612~MHz inversion dominates (Elitzur 1992). To produce 1720~MHz
inversion and conjugate 1612~MHz absorption requires the $^2\Pi_{3/2}(J=5/2)$ {-$^2\Pi_{3/2}(J=3/2)$} 
transition to be optically thick and the $^2\Pi_{1/2}(J=1/2)$ {-$^2\Pi_{3/2}(J=3/2)$} transition 
to be optically thin. The result is that 
for a column density per velocity interval just below 
$10^{15}\;$\cms \;${\rm km}^{-1}$s, 
the 1720~MHz line is inverted and the 1612~MHz line is
anti-inverted. The reverse behavior occurs for column density just above 
$10^{15}\;$\cms \;${\rm km}^{-1}$s~(van Langevelde et al. 1995).

The fitted column densities $N_{\rm OH}$ in Section~4.2 are all below $5.2\times 10^{14}$ \cms, and the line widths at all positions are greater than 1 \kms. Thus the observed physical condition do not correspond to the reverse condition mentioned above. RADEX can not reproduce the OH 1720 MHz absorption when fitting OH lines inside the boundary, as shown in Fig.~12. {There must be other pumping mechanisms, such as FIR emission from dust grains (Sivagnanam 2004), chemical pumping (Elitzur 1992), shocks (Pihlstr ̈om et al. 2008), and/or a combination of these factors to invert the satellite lines of OH. RADEX takes into account none of the above. Given the lack of protostars or \hii\ regions in the whole Taurus neighborhood and the possible existence of colliding streams seen in our data, we propose low velocity shocks as the main 
reason for the inversion of OH satellite lines.}

 \section{SUMMARY AND CONCLUSIONS}
 \label{SUMMARY}

{We have mapped a sharp boundary region of the Taurus molecular cloud in the four ground-state transitions of the hydroxyl (OH) radical with the Arecibo telescope. We then carried out a combined analysis of our OH data along with \hi, \co~J=1-0, \13co~J=1-0, with the non-LTE spectral analysis radiative transfer model, RADEX. Our main findings are the following: }
\begin{enumerate}[1.]
\item The two main lines are seen in emission in all the positions across the boundary region. However the characteristics of the two satellite lines show significant differences across the boundary region. The satellite lines at 1612 and 1720 MHz show absorption and emission respectively outside the boundary, and show emission and absorption respectively inside the boundary. 
\item Our cut perpendicular to the boundary shows that OH has two kinematic components and one component shifts from 5.3 \kms\ to 6 \kms, which {seems to indicate colliding streams} or gas flow at the boundary region. 
\item We have used a cylindrical model and RADEX to fit the OH lines to determine the physical parameters, including $T_{\rm k}$, the OPR, $N_{\rm OH}$, across the boundary. All the physical parameters can be well constrained. The excitation of OH is far from LTE. 

\item We have derived the OH abundance ($N_{\rm OH}/N_{\rm H_2}$) profile across the boundary. $N_{\rm OH}/N_{\rm H_2}$ decreases from 8$\times 10^{-7}$ to 1$\times 10^{-7}$ as $A_v$ increases from 0.4 to 2.7 mag, following an exponential law [OH]/[\h2]$=1.5\times10^{-7}+9.0\times10^{-7}\times \exp(-A_{v}/0.81)$. {OH abundance [OH]/[\h2] for moderate extinction (A$_v$~$\sim$~2) is consistent with that of PDR model. But OH at extinctions at or below 1 mag is overabundant than the prediction of PDR model by a factor of 80. The overabundance of OH is likely the result of a C-shock, which accelerates the neutral-neutral reactions.}

\item {We obtained the fraction of CO-dark molecular gas (DGF) across the boundary. The DGF decreases from 0.8 to 0.2 following a gaussian profile DGF$=0.90\times \exp(-(\frac{A_{v}-0.79}{0.71})^2)$. This trend of DGF is expected from theory. The DGF drops at low visual extinction due to the reduced abundance of \h2, and drops at high visual extinction due to the complete conversion of carbon to CO. The DGF is thus maximum in the range where \h2 has already formed but \co ~has not formed completely. This is of significance in estimating the amount of molecular gas in low extinction clouds or regions of galaxies with similar condition. }

\item {We detected the conjugate emission of OH 1612 MHz and 1720 MHz components. The complementary switching between emission and absorption of 1612 MHz and 1720 MHz, respectively, suggests that an incompletely modeled pumping mechanism must be operative, the most likely of which is C-shock. }
\end{enumerate}

\acknowledgments
\begin{acknowledgements}
This work is partly supported by the China Ministry of Science and Technology under State Key Development Program for Basic Research (973 program) No. 2012CB821802, the National Natural Science Foundation of China No. 11373038, No. 11373045, and the Strategic Priority Research Program "The Emergence of Cosmological Structures" of the Chinese Academy of Sciences, Grant No. XDB09010302. This work was carried out in part at the Jet Propulsion Laboratory, which is operated for NASA by the California Institute of Technology. {Di Li acknowledges support from the Guizhou Scientific Collaboration Program (\#20130421).} We are grateful to Carl Heiles and Z.Y. Ren for their kind and valuable advice and support. {We would like to thank the anonymous referee for the careful inspection of the manuscript and constructive comments particularly the important suggestions to add the comparison with PDR model for similar G$_0$ and n$_{\rm H}$ values to improve the quality of this study. }

\end{acknowledgements}

\end{document}